\documentclass[draft]{agujournal2019}
\usepackage{url} %this package should fix any errors with URLs in refs.
\usepackage{lineno}
\usepackage[inline]{trackchanges} %for better track changes. finalnew option will compile document with changes incorporated.
\usepackage{soul}
\usepackage[latin1]{inputenc}
\usepackage{amsmath}
\usepackage{amsfonts}
\usepackage{amssymb}
\usepackage{graphicx}
\usepackage{setspace}
\usepackage{algorithm}
\draftfalse

\journalname{Geophysical Research Letters}

\begin{document}

%% ------------------------------------------------------------------------ %%
%  Title
%
% (A title should be specific, informative, and brief. Use
% abbreviations only if they are defined in the abstract. Titles that
% start with general keywords then specific terms are optimized in
% searches)
%
%% ------------------------------------------------------------------------ %%

% Example: \title{This is a test title}

\title{\section*{Machine-Learned Preconditioners for Linear Solvers in Geophysical Fluid Flows}}

%% ------------------------------------------------------------------------ %%
%
%  AUTHORS AND AFFILIATIONS
%
%% ------------------------------------------------------------------------ %%

% Authors are individuals who have significantly contributed to the
% research and preparation of the article. Group authors are allowed, if
% each author in the group is separately identified in an appendix.)

% List authors by first name or initial followed by last name and
% separated by commas. Use \affil{} to number affiliations, and
% \thanks{} for author notes.
% Additional author notes should be indicated with \thanks{} (for
% example, for current addresses).

% Example: \authors{A. B. Author\affil{1}\thanks{Current address, Antartica}, B. C. Author\affil{2,3}, and D. E.
% Author\affil{3,4}\thanks{Also funded by Monsanto.}}

\authors{Jan Ackmann\affil{1}, Peter D. D\"uben\affil{2}, Tim N. Palmer\affil{1}, Piotr K. Smolarkiewicz\affil{3}}

\affiliation{1}{University of Oxford, Oxford, OX1 3PU, UK}
\affiliation{2}{European Centre For Medium Range Weather Forecasts, Reading, RG2 9AX, UK}
\affiliation{3}{National Center for Atmospheric Research, Boulder, CO 80026, USA}

% \affiliation{4}{Fourth Affiliation}

%\affiliation{1}{=Affiliation Address=}
%\affiliation{2}{=Affiliation Address=}
%\affiliation{3}{=Affiliation Address=}
%(repeat as many times as is necessary)

%% Corresponding Author:
% Corresponding author mailing address and e-mail address:

% (include name and email addresses of the corresponding author.  More
% than one corresponding author is allowed in this LaTeX file and for
% publication; but only one corresponding author is allowed in our
% editorial system.)

% Example: \correspondingauthor{First and Last Name}{email@address.edu}

\correspondingauthor{Jan Ackmann}{jan.ackmann@physics.ox.ac.uk}

%% Keypoints, final entry on title page.

%  List up to three key points (at least one is required)
%  Key Points summarize the main points and conclusions of the article
%  Each must be 140 characters or fewer with no special characters or punctuation and must be complete sentences

% Example:
% \begin{keypoints}
% \item	List up to three key points (at least one is required)
% \item	Key Points summarize the main points and conclusions of the article
% \item	Each must be 140 characters or fewer with no special characters or punctuation and must be complete sentences
% \end{keypoints}

\begin{keypoints}
\item The preconditioning step in linear solvers of weather and climate models can be performed using machine learning.
\item The approach can be learned from timesteps and no analytically-derived preconditioner is required as reference.
\item The machine-learned preconditioner can be interpreted in order to improve designs of conventional preconditioners.
\end{keypoints}

%% ------------------------------------------------------------------------ %%
%
%  ABSTRACT and PLAIN LANGUAGE SUMMARY
%
% A good Abstract will begin with a short description of the problem
% being addressed, briefly describe the new data or analyses, then
% briefly states the main conclusion(s) and how they are supported and
% uncertainties.

% The Plain Language Summary should be written for a broad audience,
% including journalists and the science-interested public, that will not have 
% a background in your field.
%
% A Plain Language Summary is required in GRL, JGR: Planets, JGR: Biogeosciences,
% JGR: Oceans, G-Cubed, Reviews of Geophysics, and JAMES.
% see http://sharingscience.agu.org/creating-plain-language-summary/)
%
%% ------------------------------------------------------------------------ %%

%% \begin{abstract} starts the second page

\begin{abstract}

It is tested whether machine learning methods can be used for preconditioning to increase the performance of the linear solver -- the backbone of the semi-implicit, grid-point model approach for weather and climate models.

Embedding the machine-learning method within the framework of a linear solver circumvents potential robustness issues that machine learning approaches are often criticized for, as the linear solver ensures that a sufficient, pre-set level of accuracy is reached. The approach does not require prior availability of a conventional preconditioner and is highly flexible regarding complexity and machine learning design choices.

Several machine learning methods are used to learn the optimal preconditioner for a shallow-water model with semi-implicit timestepping that is conceptually similar to more complex atmosphere models. The machine-learning preconditioner is competitive with a conventional preconditioner and provides good results even if it is used outside of the dynamical range of the training dataset.
\end{abstract}

\section*{Plain Language Summary}

The recent boom of machine-learning techniques has a huge impact on many areas of science. In this paper, we propose a new approach that is using machine learning in a part of weather and climate models called the dynamical core which is solving the discrete representation of the underlying equations of motion. In particular, we are focussing here on a model component that is required to solve an expensive linear optimisation problem each time-step. When running a model simulation, this part of the model is typically responsible for a large fraction of the computational cost.

We show how machine-learning can be used to speed-up these linear optimisation problems. We study our approach in a representative model of medium complexity that has similar properties when compared to the dynamical core of a full weather or climate model. We describe how the machine-learning approach can be applied, discuss its properties and show the performance in comparison to conventional methods. The approach is successful as it allows for stable simulations with high efficiency that are competitive with conventional model configurations.

%% ------------------------------------------------------------------------ %%
%
%  TEXT
%
%% ------------------------------------------------------------------------ %%

%%% Suggested section heads:
% \section{Introduction}
%
% The main text should start with an introduction. Except for short
% manuscripts (such as comments and replies), the text should be divided
% into sections, each with its own heading.

% Headings should be sentence fragments and do not begin with a
% lowercase letter or number. Examples of good headings are:

% \section{Materials and Methods}
% Here is text on Materials and Methods.
%
% \subsection{A descriptive heading about methods}
% More about Methods.
%
% \section{Data} (Or section title might be a descriptive heading about data)
%
% \section{Results} (Or section title might be a descriptive heading about the
% results)
%
% \section{Conclusions}

\section{Introduction}

Climate prediction models continue to show systematic deficiencies whose magnitude is comparable with the (e.g. greenhouse gas forcing) signals we seek to simulate and understand \cite{Palmer24390}. In a nonlinear system like climate, such deficiencies compromise the reliability of almost all regional outputs derived from such models. These deficiencies appear to arise from the way in which key physical processes - deep convection, orographic gravity waves and ocean mesoscale eddies in particular - are parameterised rather than modelled directly from the laws of physics. As such, an important goal for the coming years is the development of global climate models where these processes are resolved directly. However, such a goal cannot be achieved simply by relying on next-generation exascale computers - we need also to improve the computational efficiency of existing numerical codes radically. In this paper we address a central component of a broad class of climate models -- we target the dynamical core using machine learning.

 Machine learning methods show great potential for various applications across the entire workflow of weather and climate modelling including observation pre-processing, data assimilation, forecast models and post-processing. Machine learning is for example used to improve models via the development of new physical parametrisation schemes (e.g. \citeA{Schneider2018, Gentine2018}) or via emulation of existing parametrisation schemes to improve model efficiency (e.g. \citeA{Chevallier1998,Krasnopolsky2010,Rasp2018}). Other approaches aim to learn the equations of motion of the atmosphere and the ocean directly -- effectively replacing the entire dynamical core (\citeA{Dueben2018, Scher2019, Weyn2019}). Developing machine learning applications, and in particular deep learning, is highly desirable as they run very efficiently on modern supercomputers. Next-generation supercomputing hardware will be optimised (co-designed) for deep learning applications that use dense linear algebra at low numerical precision -- using 16 bits or less to represent real numbers (see for example \citeA{Kurth2018}).

This paper aims to improve computational efficiency of the dynamical core by using machine learning to develop preconditioners for linear solvers. Efficient linear solvers are essential for atmosphere and ocean models that are using implicit or semi-implicit timestepping schemes. Implicit schemes revolve around solving a problem in which the flow state at timestep ($t^{n+1}$) depends non-linearly on information from timestep ($t^{n+1}$); in contrast to explicit schemes that only use information from current/previous timesteps ($t^{n},t^{n-1},\ldots $); for a recent overview see \cite{Mengaldo2019}.
 In practice, often semi-implicit timestepping schemes are used which evaluate slow-moving parts of the equations of motion explicitly and solve implicitly for pressure to cover the fast-moving parts. As a result, (semi-)implicit methods allow for the use of much longer timesteps. For example the explicit COSMO model \cite{Fuhrer2018} uses 12 seconds, at similar resolution the fully-implicit model in \cite{Yang2016} was pushed to an extreme value of 240 seconds -- too large timesteps eventually lead to a degradation of model solutions. However, for implicit models the linear solver is responsible for the majority of the computational cost. This paper is focussing on the class of Krylov sub-space methods but the approach presented should also be relevant for multigrid-based methods, see \cite{Mueller2014, maynard2020} for recent publications on linear solvers.

To reduce the cost of the linear solver -- which typically requires a number of solver iterations for convergence -- efficient preconditioners are essential. A preconditioner directly inverts specific parts of the linear problem in order to significantly reduce the number of solver iterations. However, deriving efficient preconditioners is a difficult exercise requiring substantial research (\citeA{Mueller2014, maynard2020, Kuhnlein2019, Piotrowski2016, Dedner2016}). There have been some advances to apply machine-learning methods within the context of linear solvers. So far, work has focused on using machine learning to either select the best solver-preconditioner setup from a set of preconditioners and/or linear solvers for a given linear problem \cite{holloway2007, kuefler2008, xu2005, George2008, Yamada2018, Huang2016, PEAIRS2011}, to help improve efficiency for Block-Jacobi type preconditioners \cite{gotz2018}, to reduce the time-to-solution by interspersing linear solver iterations with neural-network based correction steps \cite{Rizzuti2019}, or to replace the linear solver entirely \cite{tompson2017accelerating,Yang2016data,Ladick2015data}.

This paper will try a fundamentally new approach by using supervised machine learning to derive the preconditioner directly. We will perform preliminary tests and train machine learning preconditioners for the application in a global shallow water model. This specific approach to preconditioning has several advantages. Machine learning methods, and deep learning methods in particular, are often criticised for potentially leading to unphysical model behaviour if the methods are used outside of the dynamic range of the training dataset, for example in a changing climate. The use of machine learning in the preconditioner is, however, not as vulnerable to these problems. If the performance of the machine-learning preconditioner degrades, the linear solver will only continue with the timestep if the error of the solution reaches a user-defined threshold. Also, for the machine-learned preconditioner the complexity of the machine learning method and its set of input variables can be freely adjusted by the user to performance requirements. This makes the machine-learned preconditioner very flexible, which is important for efficient parallel communication on supercomputers. 

Section \ref{sec:model} provides information about the model and the testcase that is used. Section \ref{sec:NN_exp} describes how machine learning is used to develop preconditioners. Section \ref{sec:results} is presenting the results including the offline performance of machine-learned preconditioners, the use of the preconditioners within free-running simulations, an investigation about the learned properties of the preconditioners, and a brief discussion of computational performance. Section \ref{sec:conclusion} provides a discussion and the conclusion.

\section{Model and Test-case}\label{sec:model}

\subsection {Shallow-Water Model}

We use an Eulerian, semi-implicit shallow-water model that is conceptionally similar to the Finite Volume Model (FVM-IFS) which is developed at the European Centre for Medium-Range Weather Forecasts as a new dynamical core for the Integrated Forecasting System (IFS; \citeA{Smolarkiewicz2016, Smolarkiewicz2019, Kuhnlein2019}). The shallow-water model is using the well-known MPDATA advection scheme \cite{Prusa2008,SzmelterSmolar10} and the shallow water equations on the sphere are discretised as defined in \cite{Smolarkiewicz1998,SzmelterSmolar10}:

\begin{equation}
\begin{aligned}\label{SWE1}
\frac{\partial G \Phi}{\partial t}+\nabla \cdot ({\bf v}\Phi) =0~,
\end{aligned}
\end{equation} 
\begin{equation}
\begin{aligned}\label{SWE2}
\frac{\partial G Q_{x}}{\partial t}+ \nabla \cdot({\bf v}Q_{x}) =  G\,R_x
\end{aligned}
\end{equation} 
\begin{equation}
\begin{aligned}\label{SWE3}
\frac{\partial G Q_{y}}{\partial t}+ \nabla \cdot({\bf v}Q_{y}) = G\,R_y~,
\end{aligned}
\end{equation}
where $\Phi$ is the fluid thickness, $Q_{x}$ and $Q_{y}$ denote the momenta in $x=\lambda$ (longitudinal) and $y=\phi$ (latitudinal) directions. $G\equiv h_{x}h_{y}$ is the Jacobian of the geospherical framework, with $h_{x}$, $h_{y}$ being the metric coefficients of the general orthogonal coordinates; here $h_{x}=a\cos(\phi)$, $h_{y}=a$ for a lat-lon grid with Earth's radius $a$. ${\bf v}$ is the advective velocity. 

The corresponding forcing terms for the momenta in equations ~\eqref{SWE2} and ~\eqref{SWE3} are:

\begin{equation}
\begin{aligned}\label{RSWE2}
R_x = -\frac{g}{h_{x}}\Phi \frac{\partial(\Phi+H_{0})}{\partial x}
+fQ_{y}+\frac{1}{G\Phi}(Q_{y}\frac{\partial h_{y}}{\partial x} - 
Q_{x}\frac{\partial h_{x}}{\partial y})Q_{y}~,
\end{aligned}
\end{equation} 
\begin{equation}
\begin{aligned}\label{RSWE3}
R_y = -\frac{g}{h_{y}}\Phi \frac{\partial(\Phi+H_{0})}{\partial y}
-fQ_{x}-\frac{1}{G\Phi}(Q_{y}\frac{\partial h_{y}}{\partial x}
-Q_{x}\frac{\partial h_{x}}{\partial y})Q_{x}~.
\end{aligned}
\end{equation} 
The terms occurring on the right-hand-sides are, from left to right, the pressure gradient, the Coriolis force, and 
the metric terms. $H_{0}$ is the topography, $g$ the gravitational acceleration and $f$ the Coriolis parameter.

Equations ~\eqref{SWE1}-\eqref{SWE3} are discretized in a semi-implicit fashion on a collocated lat-lon grid. While the explicit part of the momentum equations is evaluated via the MPDATA approach \cite{Smolarkiewicz1998}, this paper is focusing on the linear solver which is used for the implicit part of the time integration. The linear problem to be solved for $\Phi^{n+1}$  originates from inserting the trapezoidal integrals of the momentum equations ~\eqref{SWE2} and ~\eqref{SWE3} into the trapezoidal integral of equation ~\eqref{SWE1}. It can be symbolically written as: 

\begin{equation}\label{Elliptic}
\mathcal{L}\left(\Phi^{n+1}\right)-\mathcal{R}=0.
\end{equation} 
For illustration, $\mathcal{R}$ incorporates the explicit parts of the time integration that were already computed outside of the solver, while $\mathcal{L}$ represents the discretised implicit terms. Here, $\mathcal{L}$ is a linear operator which is negative-definite but not self-adjoint. $\mathcal{L}$ has the form of a generalized Laplacian:

\begin{equation}\label{L_Operator}
\mathcal{L}\left(\Phi\right):=\sum^{2}_{I=1}\frac{\partial}{\partial x^{I}}
\left(\sum^{2}_{J=1}A^{IJ}\frac{\partial \Phi}{\partial x^{J}} + B^{I}\Phi\right)-\Phi~,
\end{equation} 
with the six coefficient fields: $A^{11}$ (zonal direction), $A^{12},A^{21}$ (cross-derivative terms), $A^{22}$ (meridional direction), $B^1$ and $B^2$.

After spatially discretising the linear problem ~\eqref{Elliptic}, it can be solved using a linear solver, here the preconditioned generalized conjugated residual method (GCR; see \citeA{smolarkiewicz2000variational, eisenstat1983variational} and flowchart in supporting information), a Krylov subspace method that iteratively minimizes the Euclidean norm of the residual vector $r_{\nu}$ at each solver iteration $\nu$: 

\begin{equation}\label{Residual}
r_\nu = \mathcal{L}\left(\Phi_\nu\right)-{\mathcal R}~.
\end{equation} 
The solver is iterated until the infinity norm of the residual is found to be smaller than an $\epsilon$ value that is adjusted to application needs $\frac{\parallel r_{\nu} \parallel_\infty\, }{\parallel r_{0} \parallel_\infty\, } \leq\, \epsilon\,$ (results are qualitatively the same if using the Euclidean norm instead; not shown). If the solver is converged after $N$ iterations, we set $\Phi^{n+1}:=\Phi_{N}$. 

We use an 'implicit Richardson' preconditioner $\mathcal{P}$ as reference to allow for a qualitative comparison between conventional and machine learning approaches (see supporting information), that is based on performing implicit Richardson iterations in zonal direction to diminish the effects of grid-convergence near the poles. This approach is equivalent to the well-known treatment of the vertical dimension with a tridiagonal solver, see \cite{Mueller2014, maynard2020}. The preconditioner has been successfully tested for shallow-water test-cases from the Williamson test-suite \cite{Williamson} and allowed for solver speed-ups of factor $3$ to $10$ (not shown here).

%\begin{figure}[!htb]
%\centering
%  \includegraphics[width=1.0\textwidth, trim=0cm 5.5cm 0cm 5cm, %clip]{plots/NN_architecture.pdf}
%\caption{Schematic of a feed-forward deep neural network of %artificial neurons with input vector $x$ and output $\mathcal{NN}%(x)$.}
%\label{fig:NN_schem}
%\end{figure}
\subsection {Test-case}

\begin{figure}[!htb]

  \includegraphics[trim=3.4cm 18.5cm 0cm 2cm, clip]{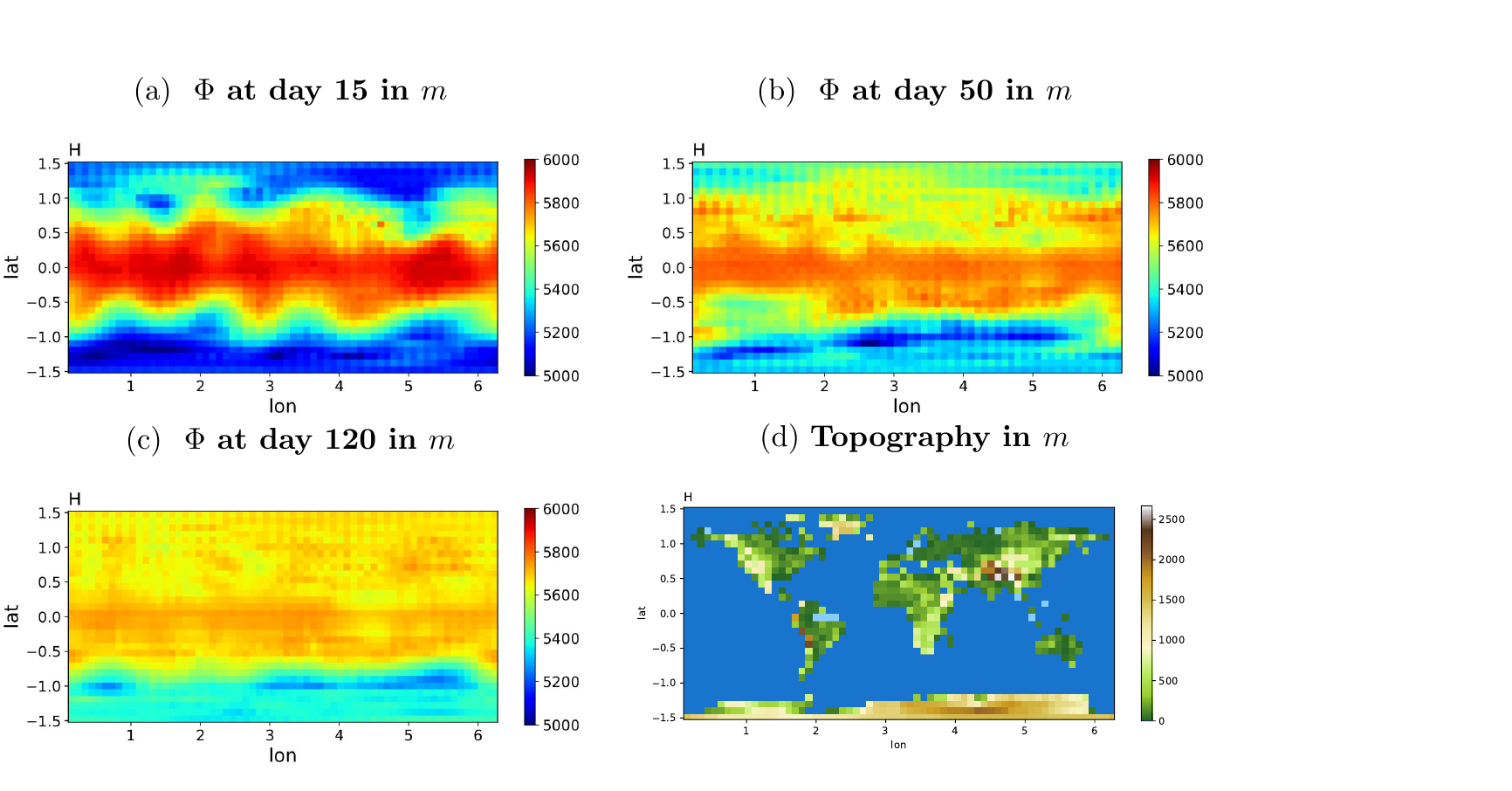}

\caption{Snapshots of fluid thickness $\Phi$ in [m] after 15 (a), 50 (b), 120 (c) simulation days, as well as the model topography in [m] (d). Latitude and longitude are in radians.}
\label{fig:fields}

\end{figure}

We apply the model to the zonal geostrophic flow test-case as described in \cite{Williamson} with the parameters $\alpha=0$, $u_{0}=20\frac{m}{s}$, and $h_{0}=5960m$. However, to increase the complexity of the test-case, we are adding real-world topography that is based on the ETOPO5 dataset \cite{ETOPO5}. The original topography is limited to positive values only, and scaled by a factor $0.5$ to ensure that topography is covered by the fluid at all times. We run the model at a $5.6^{\circ}$ model resolution ($64$$\times$$32$ grid-points). The timestep length is chosen to be $240 s$, which satisfies the Courant number requirement of the chosen discretisation. We choose a comparably small $\epsilon$ value of $1 \cdot 10^{-10}$ to study convergence of the solver over a wide dynamic range. The model is run for 120 days. Figure \ref{fig:fields} is showing snapshots of the model state. Due to interactions with the topography, the initial zonal jet structure is decaying, leading to an abundance of different flow states and scales.

\subsection{The Machine-Learned Preconditioner}
\label{sec:NN_exp}

It is the task of a preconditioner $\mathcal{P}$ to reduce the overall workload required by the solver. The preconditioner achieves this by providing an estimate of the solution error $\mathcal{P}^{-1}(r_{\nu})\approx \mathcal{L}^{-1}(r_{\nu})=\Phi^{n+1}-\Phi_{\nu}$, the increment to the fluid thickness $\Phi_{\nu}$ that is required to reach the next timestep $\Phi^{n+1}$. We therefore train our machine-learned preconditioners to predict an estimate $\Delta \tilde{\Phi}$ of the required increment $\Delta \Phi:=\Phi^{n+1}-\Phi^{n}$. 

As the machine-learning tool of choice, we use fully-connected, feed-forward, neural networks. For these networks, the value of the $i$-th neuron of the $k$-th hidden layer $y^{(k)}_{i}$ is the result of applying an activation function $\phi$ to the weighted sum of outputs $y^{(k-1)}_{i}$ from the previous layer $k-1$: 

$$y^{(k)}_{i}=\phi\left(\sum_{j=0}^{m_{k-1}}w^{(k,\: i)}_{j}y^{(k-1)}_{j}+b_{k}\right),$$ 
where $m_{k-1}$ is the number of neurons in layer $k-1$, $w^{(k,\: i)}$ is the vector of weights of the $i$-th neuron of layer $k$, and $b_{k}$ is a bias term. For the hidden layers, the ReLU activation function $\phi_{ReLU}(x):=max(0,x)$ provided the best results. We use a linear activation function for the output layer.

We set up the neural network to predict a single grid-point value of $\Delta \tilde{\Phi}$ at a time. The input is based on grid stencils of the $6$ coefficients ($A^{ij}$ and $B^i$) of the linear operator $\mathcal{L}$ (which are constant throughout the timestep) plus the residual $r_{\nu}$.  
We normalize the six coefficient fields to stay in the interval $\left[-0.5,0.5\right]$. Motivated by the linearity of operator $\mathcal{L}$, the input residual values $r_{\nu}$ and output values $\Delta \tilde{\Phi}$ are rescaled via division by $2\parallel r_{\nu} \parallel_\infty$. This rescaling aims at making the machine-learned preconditioner invariant to the shrinking dynamical range of residual values $r_{\nu}$ over subsequent solver iterations. 
%\begin{table}\caption{Chosen values for input normalization of the linear operator coefficients}
%\begin{center}
%\begin{tabular}{|c|c|c|c|c|c|c|}\hline
%    & $A^{11}$              & $A^{12}$& $A^{21}$& $A^{22}$& $B^{1}$& $B^{2}$\\\hline
%  Mean & 1.176e+11     &  0              &            0   & 2.574e+10    &       0          &  0 \\\hline
%	Max-Min & 8.293e+11 &  1.465e+9 & 1.465e+9 &   4.214e+10 &   1.052e+10 & 5.272e+9\\\hline
% \end{tabular}
%\end{center}\label{tab: Tabelle1}
%\end{table}

Reminiscent of local approximate inverse preconditioners \cite{SMITH199238}, we use '$3\times3$' and '$5\times5$' stencils of the input fields to predict $\Delta \tilde{\Phi}$ for the grid-point in the centre. A $5\times5$ stencil that predicts $\Delta \tilde{\Phi}$ at grid-point $i,j$ uses the following set of grid indices

$$\left\{ (k,l): k=i-2,\ldots,i+2; \; l=j-2,\ldots,j+2 \right\}$$  for all input fields. Near the poles, the stencil is completed by continuing the meridians over the poles -- in accordance with the underlying discretisation.

For the training and validation sets, we use data of the first solver iteration from the 120 day reference simulation. We train a neural network for each latitudinal band separately. To make sure that the training and validation data-sets are sufficiently independent, we built them as follows: The initial $14$ days of model integration time are omitted to avoid potential shocks in the data from initialization. Afterwards, the data is split into cycles of three weeks of integration time. For each of the three week cycles, we use all timesteps from days $1-14$ for training, omit day $15$, use all timesteps from days $16-20$ for validation, and omit day $21$. This results in $1.6 \cdot 10^{6}$ training samples and $6 \cdot 10^{5}$ validation samples for each latitude.

Neural networks of different sizes are trained and tested. The neural network size ranges from 5 hidden layers and 200 neurons per layer (L5N200) to 1 hidden layer with 5 neurons (L1N5). However, we also tested a linear regression model (L0N0). All neural networks are implemented in Keras \cite{chollet2015keras} and trained using the Adam stochastic optimization \cite{kingma2014adam, j_2018on}. The loss function is the mean-squared-error (MSE) metric. The neural networks are trained for at least 50 epochs, with a batch size of 32.
\section{Results}
\label{sec:results}

\subsection{Offline Performance of the Machine-Learned Preconditioners}

\begin{figure}[!htb]

  \includegraphics[trim=3.4cm 9.2cm 0cm 2.2cm, clip]{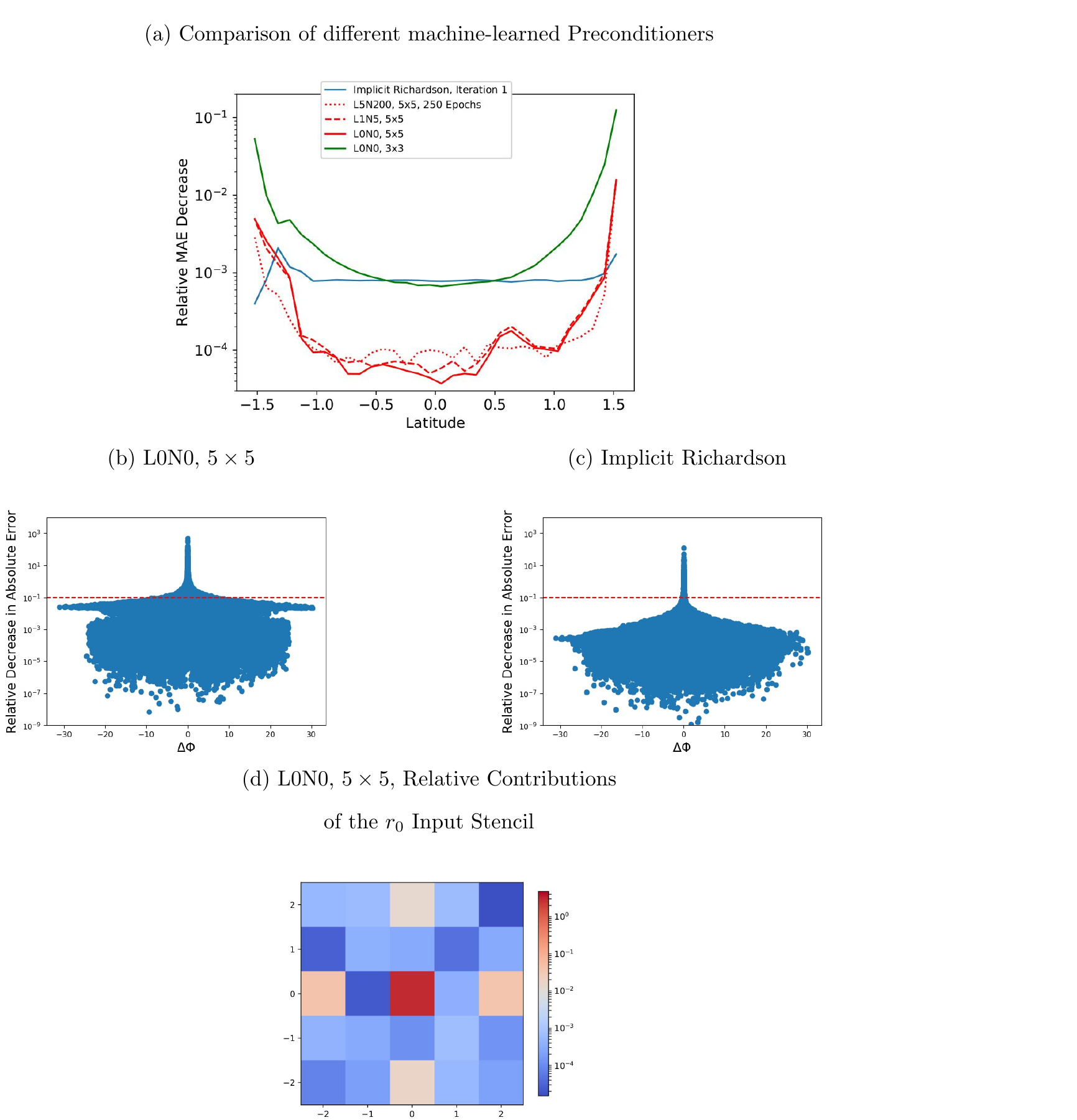}

\caption{a) Relative decrease in Mean Absolute Error (MAE) as a function of latitude for the validation set for the implicit Richardson Preconditioner in comparison to neural networks of various sizes and input stencil widths. b) and c) show the absolute values of $\frac{\Delta \tilde{\Phi}-\Delta \Phi}{\Delta \Phi}$ as a function of $\Delta \Phi$ for the zonal band closest to the South pole for b) the L0N0, $5\times 5$ preconditioner, and c) the implicit Richardson preconditioner. d) shows the mean absolute of relative contributions of the residual $r_{0}$ input stencil towards the error prediction $\Delta \tilde{\Phi}$ of the L0N0, $5\times 5$ preconditioner at latitude $\phi=1.03$ ($59^{\circ} N$).}
\label{fig:1}
\end{figure}

To get a first impression about the quality of results, we compare the relative decrease in Mean Absolute Error (MAE) at each latitude for the first iteration of the solver between the simulation using the implicit Richardson and the Neural Network preconditioner in Figure \ref{fig:1}(a). 
Lower values mean better performance, i.e. higher error reduction.

In the first solver iteration, the implicit Richardson preconditioner manages a relative decrease in MAE of 3 orders of magnitude. As expected, by design this type of preconditioner does a good job at providing stable performance near the poles despite the poleward grid convergence.

In comparison, the $5\times5$ stencil machine-learned preconditioners achieve a relative decrease in MAE of $2\cdot 10^{-2}$ and $5\cdot 10^{-3}$ near the North- and South-poles respecitively. The machine-learning preconditioners performs even better equatorwards where they achieve a relative reduction in MAE of up to $5 \cdot 10^{-5}$. Surprisingly, the size of the neural network has no significant impact. The L5N200 preconditioner is not consistently better than the L0N0 preconditioner. At the same time, the L5N200 preconditioner needed an increased number of 250 epochs for convergence. Using other activation functions for the L5N200 neural network such as $tanh$ does not improve results (not shown here).
 
As the L0N0 preconditioner will be the cheapest machine-learning preconditioner, the L0N0 with a $5\times 5$ stencil can be considered the most promising option. To complete our analysis, we train a L0N0 preconditioner on the smaller $3\times 3$ stencil. Although behaving qualitatively similar to the L0N0, $5\times 5$ preconditioner, it consistently performs worse by about one order of magnitude at all latitudes. 

To better understand the structure of the relative errors and whether the machine-learning preconditioners behave robustly, we further analyze the behavior of the L0N0, $5 \times 5$ preconditioner. We show the absolute values of relative error $\frac{\Delta \tilde{\Phi}-\Delta \Phi}{\Delta \Phi}$ as a function of $\Delta \Phi$ for the first solver iteration in Figure \ref{fig:1} (b) and (c) for the L0N0, $5 \times 5$ and the implicit Richardson preconditioner. The data shown is for the zonal band closest to the South Pole, the qualitative results are the same for the other latitudes (not shown here).

%\begin{figure}[!htb] \centering
%\caption{}
%\label{fig:convergence_Moun_res1_2}
%\end{figure}

The L0N0, $5\times 5$ preconditioner and the implicit Richardson preconditioner share the same qualitative behaviour. For a large range of $\Delta \Phi$ values, both preconditioners are good and robust predictors and reduce $\Delta \Phi$ by at least one order of magnitude (red, horizontal line). The reduction per preconditioner application is larger for the implicit Richardson preconditioner. Only for very small values of $\Delta \Phi$, we find that applying the preconditioner actually increases the error. This occurs for both preconditioners and is simply a reminder that the task of preconditioning is inherently a problem of making a prediction with incomplete information.

\subsection{Online Performance of the Machine-learned Preconditioner}\label{ML_Precon_Online}

\begin{figure}[!htb]

  \includegraphics[trim=3.4cm 14cm 0cm 2.2cm, clip]{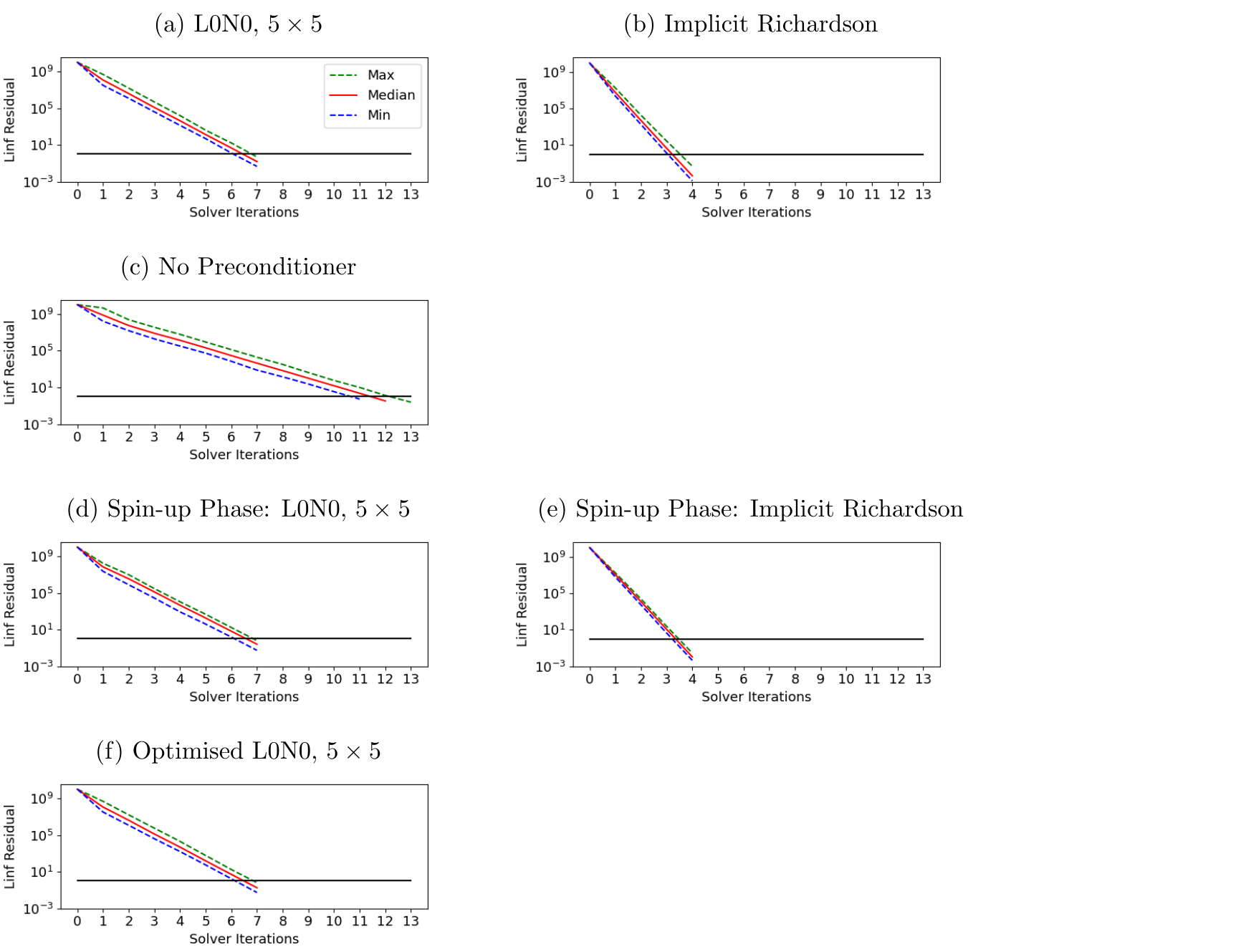}

\caption{Convergence rate analysis of the linear solvers for the implicit Richardson preconditioner and the L0N0, $5$$\times$$5$ preconditioner. The shown maxima, minima, and median values of the residual norms $\parallel$$r_{\nu}$$\parallel_\infty$ are normalized by $\epsilon$$\parallel$$r_{0}$$\parallel_\infty$. The convergence rate of the L0N0, $5$$\times$$5$ preconditioner for days 15-120 is shown in (a), this simulation is run from perturbed initial conditions (initial fields are randomly perturbed by 5$\%$). Respective convergence rates using the implicit Richardson preconditioner and no preconditioner are shown in (b) and (c), respectively. For days 0-15, the performance of the L0N0, $5$$\times$$5$ preconditioner is shown in (d), with the respective implicit Richardson reference given in (e). In (f), we show the same information as in a) for the optimised version of the L0N0, $5$$\times$$5$ preconditioner (see Section \ref{Contributions}).}
\label{fig:convergence}

\end{figure}

We now use the L0N0, $5 \times 5$ preconditioner within free-running simulations of the shallow-water model.
For our first convergence test, we use the L0N0, $5\times 5$ preconditioner and run a simulation until day 120. To rule out the possibility that our machine-learned preconditioner is only valid for one specific flow trajectory (assuming that the pre-chosen solver accuracy is so high that it results in the same trajectory), we show the L0N0, $5 \times 5$ preconditioner results from a simulation where initial conditions for the momentum $Q_{x}$ are stochastically perturbed by 5$\%$ of their value. The perturbed and unperturbed runs slowly decorrelate, the correlation coefficient for $Q_{y}$ after 50 days is down to 0.44. Note that, the L0N0, $5 \times 5$ preconditioner is used for all solver iterations and we thus go beyond the training and validation data that only used data from the first solver iteration. The convergence rate using the L0N0, $5\times 5$  preconditioner (Figure \ref{fig:convergence} (a)) is almost doubled compared to running the model without a preconditioner (Figure \ref{fig:convergence} (c)) for all iterations but almost halved when compared to the implicit Richardson preconditioner (Figure \ref{fig:convergence} (b)). 

This convergence behavior is also found for the first 15 days of integration time of the test-case -- the initial spin-up phase -- which lies outside of the training data, see Figures \ref{fig:convergence} (d) and \ref{fig:convergence} (e). 

The L0N0, $5 \times 5$ preconditioner thus consistently increases the convergence rate of the linear solver in different scenarios that were not part of the training data.

\subsubsection{Interpretability of the Machine-learned Preconditioner}\label{Contributions}

Because of the low complexity of the L0N0, $5 \times 5$ preconditioner -- please note that L0N0 is equivalent to the use of linear regression -- we can dissect how each input contributes towards the final prediction $\Delta \tilde{\Phi}$. We show the mean absolute relative contribution of the residuals $r_{0}$ input stencil in Figure \ref{fig:1} (d). Here, we only show the contribution for latitude $59^{\circ} N$. However, results are qualitatively the same for the other latitudes. 

In the contributions of the residuals $r_{0}$ (Figure \ref{fig:1} (d)), the underlying stencil of the generalized Laplace operator $\mathcal{L}$ in equation ~\eqref{L_Operator} shines through. We see 5 dominant values: two in zonal direction, two in meridional direction, and the center-most grid-point. The center-most grid-point value has the largest contribution and represents the inversion of the main diagonal terms.

The contributions of the coefficients $A^{ij}$ and $B^i$ (see supporting information) are found potentially negligible as they contribute less than 2.5$\%$ to the total of the final prediction values of $\Delta \tilde{\Phi}$.

Thus, we train an optimised version of the L0N0, $5 \times 5$  preconditioner that only uses the residuals $r_{\nu}$ as inputs. As most of these residual inputs within the stencil have negligible contributions as well, we further restrict the L0N0, $5 \times 5$ preconditioner input and use only data from local stencil coordinates $(0,2)$, $(-2,0)$, $(0,0)$, $(2,0)$, and $(0,-2)$ (with $(0,0)$ being the centre of the stencil). The resulting model simulation with the ``optimised'' L0N0 preconditioner which is using only five input values performs as well as the standard version, see Figure \ref{fig:convergence} (f). 

\subsection{Performance Estimates}

To compare the overall efficiency of the solvers, the computational overhead of the preconditioners needs to be taken into account.
In a sequential setting and small problem sizes, a simple analysis of the required floating-point operations is a good performance model for the preconditioned linear solvers. 

The optimised L0N0, $5\times 5$ preconditioner requires almost twice as many solver iterations for the same accuracy compared to the implicit Richardson preconditioner. However, each application of the machine-learned preconditioner (10 floating-point operations per grid-point) is 4 times cheaper than the implicit Richardson preconditioner (38 floating-point operations). In summary, including the rest of the elliptic solver steps + solver and preconditioner initialisation, both preconditioners result in the same computational saving of about 30 percent over the unpreconditioned elliptic solver.

\section{Discussion and Conclusion}
\label{sec:conclusion}
We show a proof-of-concept for the derivation of a machine-learned preconditioner in a representative shallow-water model. The machine-learned preconditioner performs equally well as the implicit Richardson preconditioner that was used as a reference. This is a positive result that shows the great potential of our approach. In fact, it is rather surprising that a very simple machine learning solution -- which is basically using simple linear regression -- is already sufficient to perform so efficiently for a two-dimensional fluid problem. 

The performance estimates that were presented in this paper will likely not hold for larger problem sizes and supercomputing environments. Here, performance will mostly be limited by data movement rather than floating point arithmetic. However, machine-learned preconditioners have clear advantages when compared with a conventional preconditioner as they are based on local grid stencils whose shape and size can be flexibly chosen and optimised to yield the best performance on any given hardware system. Furthermore, deep learning applications can make very efficient use of modern hardware since they are based on dense linear algebra. The use of reduced numerical precision may help to further reduce the cost of machine-learned preconditioners.

The test configuration discussed in this paper is still simple when compared to the complexity of the task to develop efficient preconditioners for high-resolution, three-dimensional atmosphere or ocean models, mainly due to the stiffness caused by vertical grid spacing of as little as a couple of meters close to the surface. However, as the power of deep learning to learn complex non-linear relationships with complex neural networks was not even required for the shallow water model, we are optimistic that machine learning preconditioners will perform well for three-dimensional models. Also, our approach might not be limited just to preconditioners because smoothers for multigrid solvers might be derived in a very similar fashion.

\appendix

\acknowledgments
Jan Ackmann and Tim Palmer were funded via the European Research Council project ITHACA (grant No. 741112). Peter Dueben gratefully acknowledges funding from the Royal Society for his University Research Fellowship. Peter Dueben and Piotr Smolarkiewicz received funding from the ESiWACE and ESiWACE2 projects. ESiWACE and ESiWACE2 have received funding from the European Union's Horizon 2020 research and innovation programme under grant agreement No 675191 and 823988. NCAR is sponsored by the National Science Foundation. We would like to thank Christian Kuehnlein for many helpful discussions. All source code can be found in repository:
https://github.com/JanAckmann/MLPrecon.

%% ------------------------------------------------------------------------ %%
%% References and Citations

%%%%%%%%%%%%%%%%%%%%%%%%%%%%%%%%%%%%%%%%%%%%%%%
%
% \bibliography{<name of your .bib file>} don't specify the file extension
%
% don't specify bibliographystyle
%%%%%%%%%%%%%%%%%%%%%%%%%%%%%%%%%%%%%%%%%%%%%%%

%\bibliography{science}
\bibliography{references_Peter}
%\include{./si_ML_precon_arxive2}

%Reference citation instructions and examples:
%
% Please use ONLY \cite and \citeA for reference citations.
% \cite for parenthetical references
% ...as shown in recent studies (Simpson et al., 2019)
% \citeA for in-text citations
% ...Simpson et al. (2019) have shown...
%
%
%...as shown by \citeA{jskilby}.
%...as shown by \citeA{lewin76}, \citeA{carson86}, \citeA{bartoldy02}, and \citeA{rinaldi03}.
%...has been shown \cite{jskilbye}.
%...has been shown \cite{lewin76,carson86,bartoldy02,rinaldi03}.
%... \cite <i.e.>[]{lewin76,carson86,bartoldy02,rinaldi03}.
%...has been shown by \cite <e.g.,>[and others]{lewin76}.
%
% apacite uses < > for prenotes and [ ] for postnotes
% DO NOT use other cite commands (e.g., \citet, \citep, \citeyear, \nocite, \citealp, etc.).
%

\end{document}

%% ------------------------------------------------------------------------ %%
%
%  TITLE
%
%% ------------------------------------------------------------------------ %%

%\includegraphics{agu_pubart-white_reduced.eps}

\title{Supporting Information for "Machine-Learned Preconditioners for Linear Solvers in Geophysical Fluid Flows"}
%
% e.g., \title{Supporting Information for "Terrestrial ring current:
% Origin, formation, and decay $\alpha\beta\Gamma\Delta$"}
%
%DOI: 10.1002/%insert paper number here%

%% ------------------------------------------------------------------------ %%
%
%  AUTHORS AND AFFILIATIONS
%
%% ------------------------------------------------------------------------ %%

% List authors by first name or initial followed by last name and
% separated by commas. Use \affil{} to number affiliations, and
% \thanks{} for author notes.
% Additional author notes should be indicated with \thanks{} (for
% example, for current addresses).

% Example: \authors{A. B. Author\affil{1}\thanks{Current address, Antartica}, B. C. Author\affil{2,3}, and D. E.
% Author\affil{3,4}\thanks{Also funded by Monsanto.}}

\authors{Jan Ackmann\affil{1}, Peter D. D\"uben\affil{2}, Tim N. Palmer\affil{1}, Piotr K. Smolarkiewicz\affil{3}}

\affiliation{1}{University of Oxford, Oxford, OX1 3PU, UK}
\affiliation{2}{European Centre For Medium Range Weather Forecasts, Reading, RG2 9AX, UK}
\affiliation{3}{National Center for Atmospheric Research, Boulder, CO 80026, USA}
%(repeat as many times as is necessary)

%% ------------------------------------------------------------------------ %%
%
%  BEGIN ARTICLE
%
%% ------------------------------------------------------------------------ %%

% The body of the article must start with a \begin{article} command
%
% \end{article} must follow the references section, before the figures
%  and tables.

\begin{article}

%% ------------------------------------------------------------------------ %%
%
%  TEXT
%
%% ------------------------------------------------------------------------ %%

\noindent\textbf{Contents of this file}
%%%Remove or add items as needed%%%
\begin{enumerate}
\item Text S1
\item Figure S1 to S2
%if Tables are larger than 1 page, upload as separate excel file
\end{enumerate}

\begin{figure}[!htb]
\centering
  \includegraphics[width=0.9\textwidth, trim=0cm 0cm 19.5cm 0cm, clip]{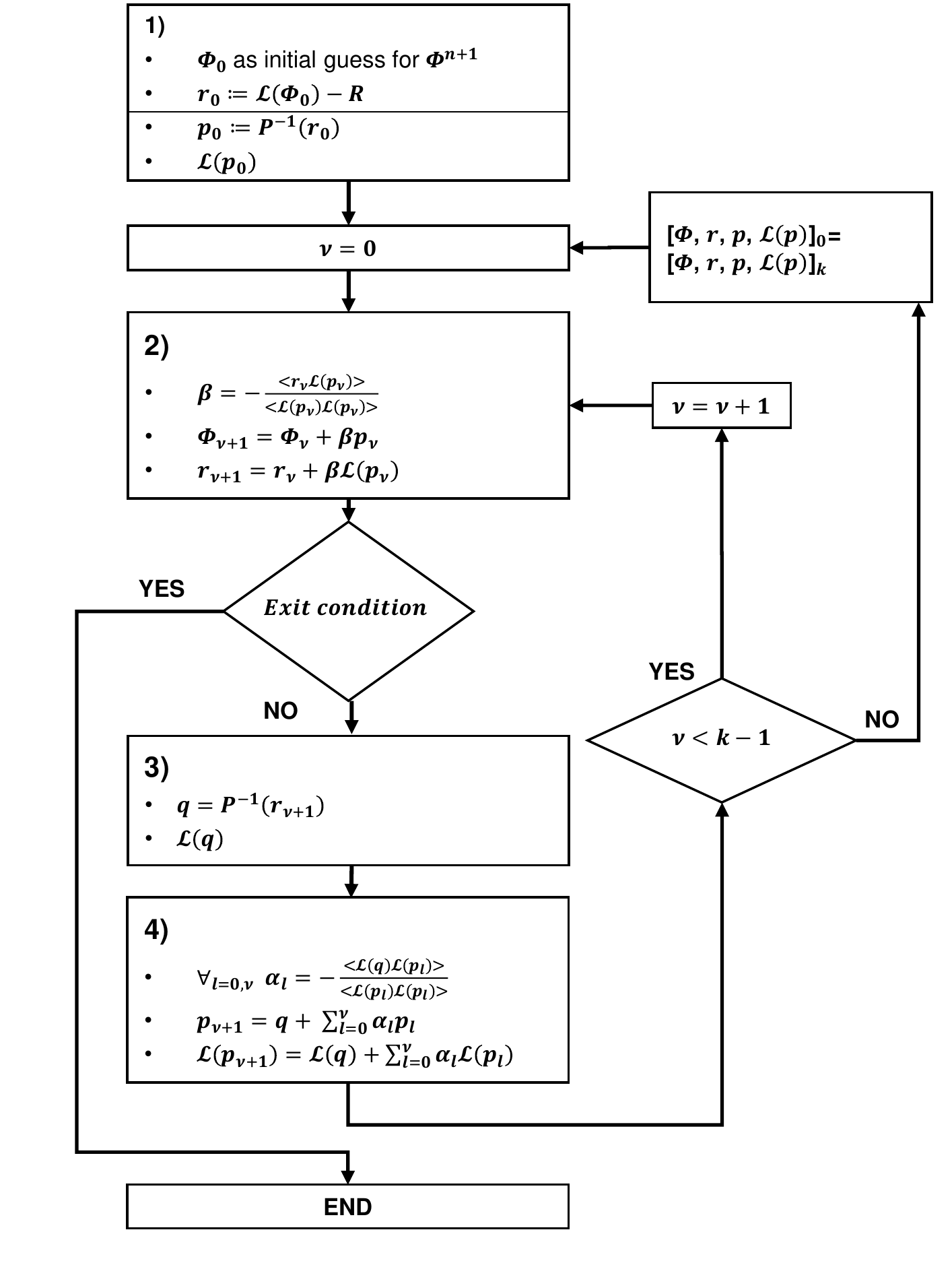}

\caption{We show a flow chart of the generalized Conjugated Residual method GCR($k$) that is implemented for this work (here $k=1$)}
\label{fig:solver}
\end{figure}

\begin{figure}[!htb]

\minipage{0.3\textwidth}(a) \centering
       	{$A^{11}$}\par\medskip
  \includegraphics[width=1.0\linewidth, trim=0cm 0cm 0cm 0cm, clip]{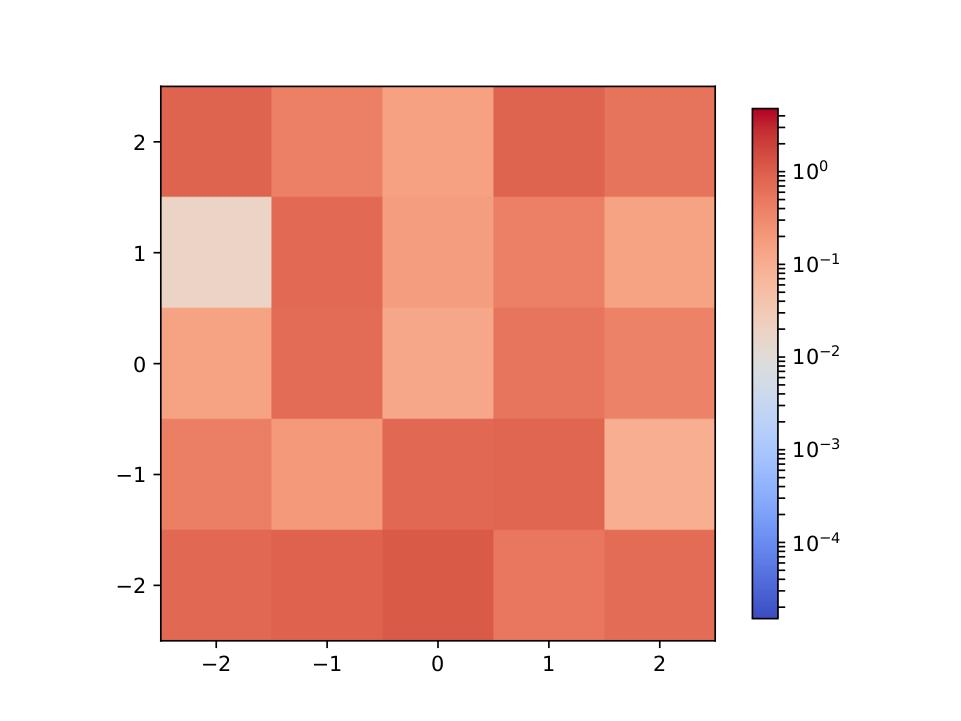}
\endminipage
\minipage{0.3\textwidth}(b) \centering
       	{$A^{12}$}\par\medskip
  \includegraphics[width=1.0\linewidth, trim=0cm 0cm 0cm 0cm, clip]{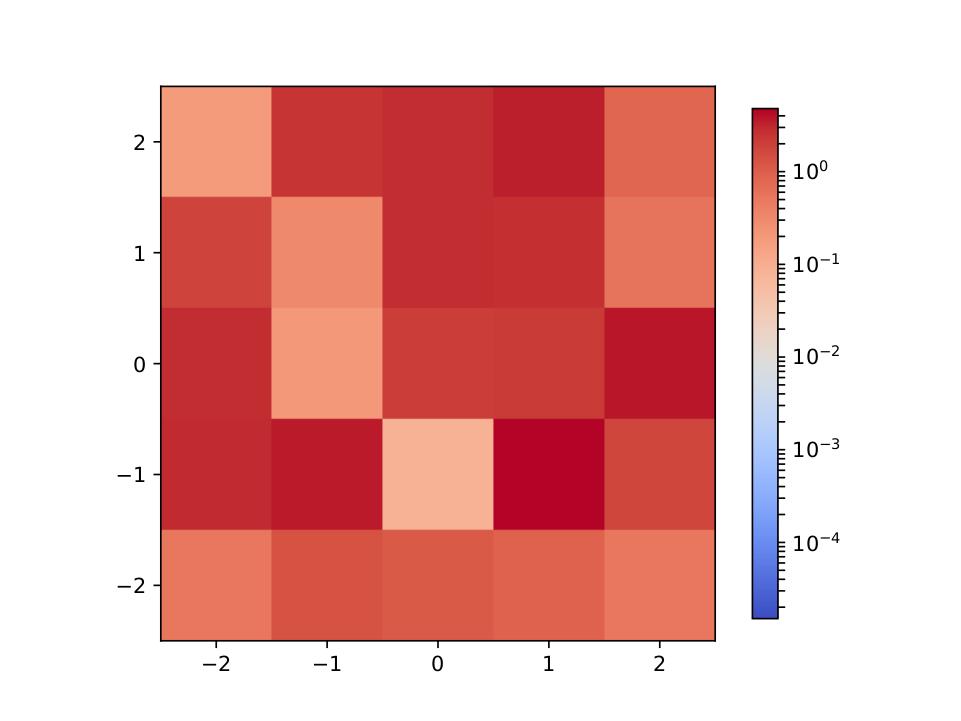}
\endminipage
\minipage{0.3\textwidth}(e) \centering
       	{$B^{1}$}\par\medskip
  \includegraphics[width=1.0\linewidth, trim=0cm 0cm 0cm 0cm, clip]{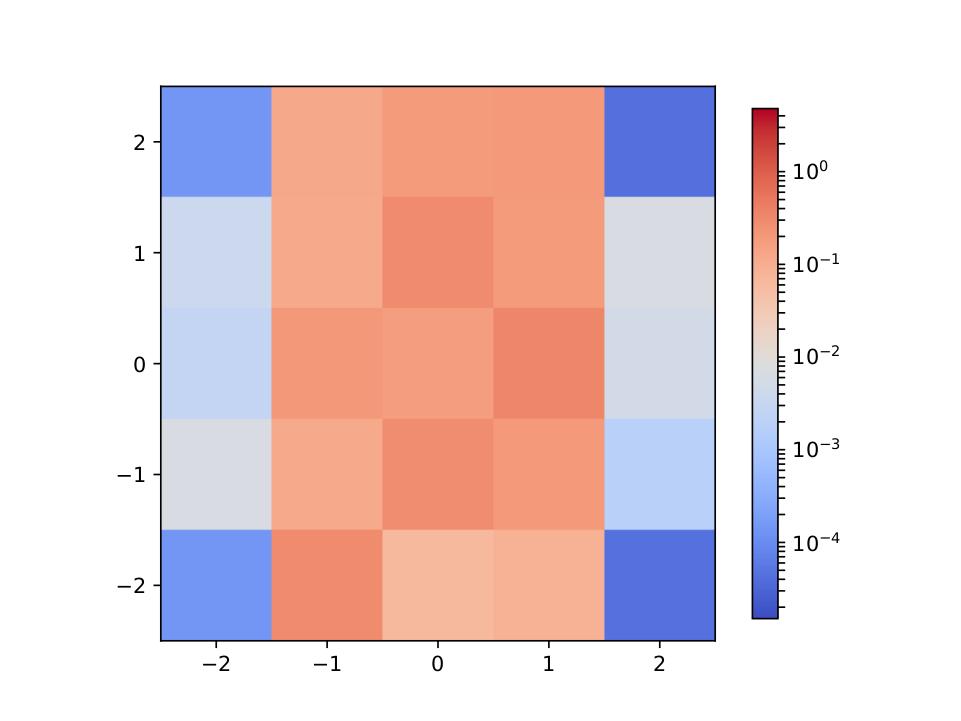}
\endminipage\hfill

\minipage{0.3\textwidth}(c) \centering
       	{$A^{21}$}\par\medskip
  \includegraphics[width=1.0\linewidth, trim=0cm 0cm 0cm 0cm, clip]{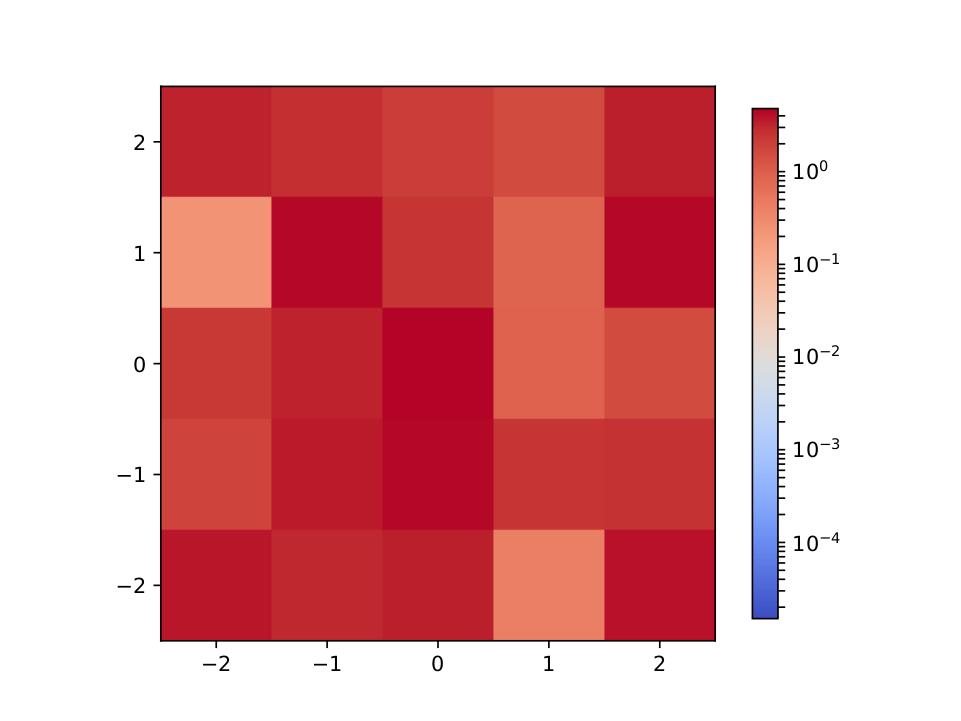}
\endminipage
\minipage{0.3\textwidth}(d) \centering
       	{$A^{22}$}\par\medskip
  \includegraphics[width=1.0\linewidth, trim=0cm 0cm 0cm 0cm, clip]{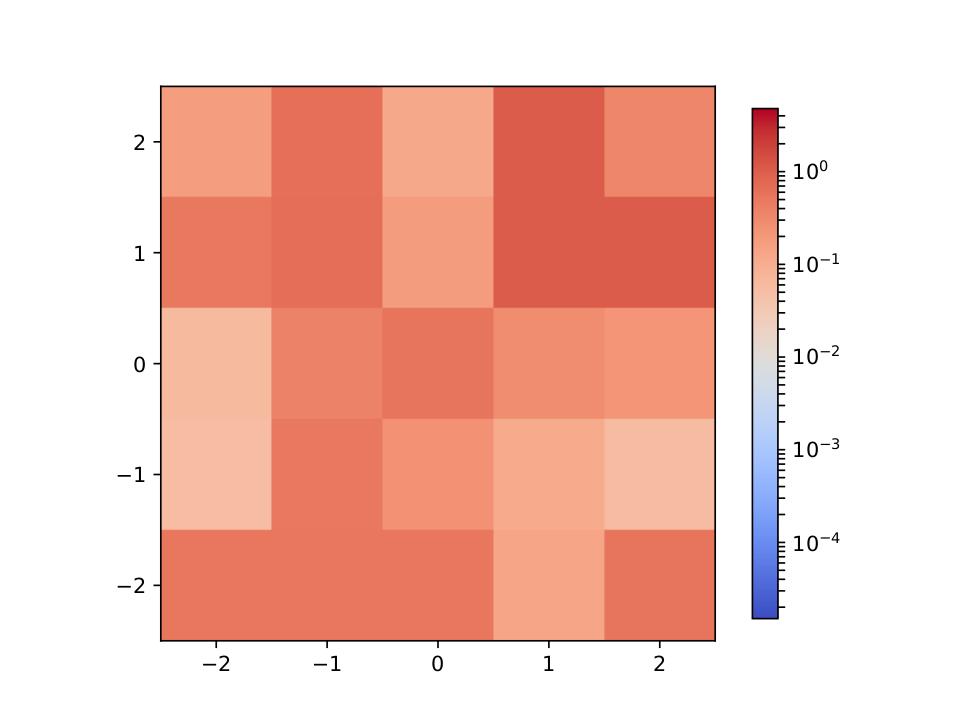} 
\endminipage
\minipage{0.3\textwidth}(f) \centering
       	{$B^{2}$}\par\medskip
  \includegraphics[width=1.0\linewidth, trim=0cm 0cm 0cm 0cm, clip]{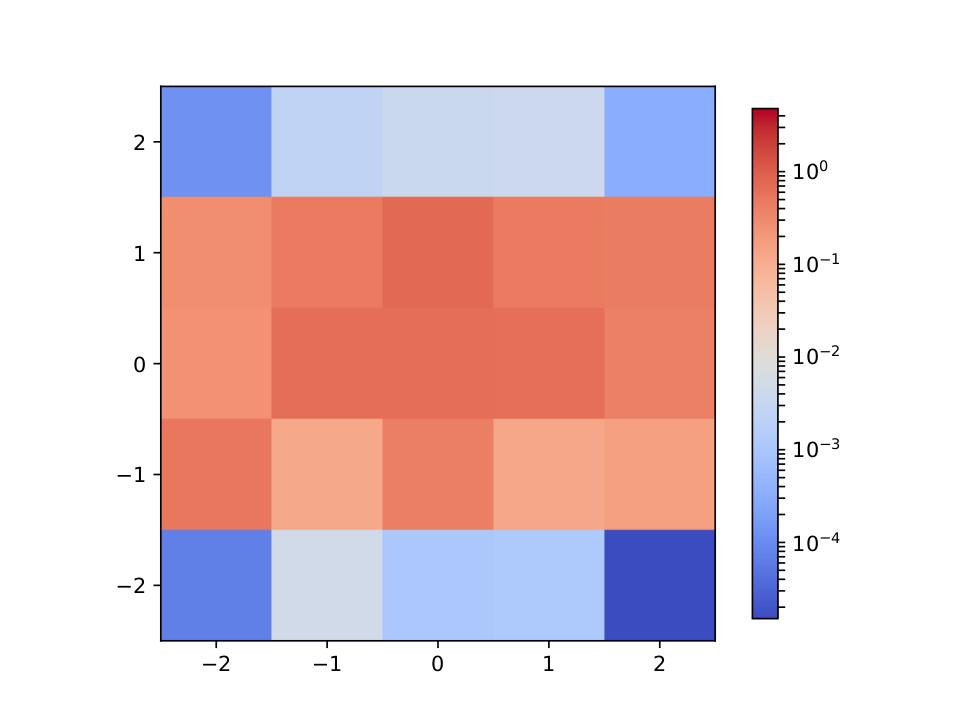} 
\endminipage\hfill

\caption{Mean absolute relative contribution of input fields towards the error prediction $\Delta \tilde{\Phi}$ for latitude $\phi=1.03$ ($59^{\circ} N$). The plots show the $5 \times 5$ stencil of the two-dimensional fields to precondition the grid-point in the centre. a), b), c), and d) show the contribution of the components of $A^{11}$, $A^{12}$,$A^{21}$, and $A^{22}$ respectively. e), f) show the contribution of the components of $B^{1}$, and $B^{2}$.}

\label{fig:contri}
\end{figure}

\noindent\textbf{Introduction}
In this supplementary information file, the reference implicit Richardson preconditioner is described in more detail in Text S1. Figure \ref{fig:solver} is a flowchart of the Generalized Conjugated Residual method (GCR) that was used in this publication. Figure \ref{fig:contri} provides additional information on the contributions from the input stencils of the linear operator for the L0N0, $5\times 5$ preconditioner.  
%Type or paste your text here. The introduction gives a brief overview of the supporting information. You should include information %about as many of the following as possible (when appropriate):
% 1. a general overview of the kind of data files;
% 2. information about when and how the data were collected or created;
% 3. a general description of processing steps used;
% 4. any known imperfections or anomalies in the data.

%\clearpage

%Delete all unused file types below. Copy/paste for multiples of each file type as needed.
\noindent\textbf{Text S1.}
Implicit Richardson based Preconditioner\newline

The implicit Richardson preconditioner obtains an estimate of the solution error $\mathcal{P}^{-1}(r_{\nu+1})\approx \mathcal{L}^{-1}(r_{\nu+1})$ by means of a stationary iteration, indexed with $\mu$ and initialized with $q_{0}=0$ --- best described as a semi-implicit Richardson scheme. For this, the operator is split into two parts. The first part combines the second-order zonal derivative term and the Helmholtz term, $\mathcal{P}^{\mathcal{Z}}+\mathcal{P}^{\mathcal{H}}$. The second part, denoted by $\mathcal{P}^{\mathcal{M}}$, consists of the second-order meridional derivative term. In the semi-implicit Richardson scheme, the first part is then taken at iteration $\mu+1$ while the second part is lagged behind. This results in a tridiagonal problem
\begin{equation}\label{P_Operator}
\left[I-\eta \mathcal{P}^{\mathcal{Z}}-\eta\mathcal{P}^{\mathcal{H}}\right]q_{\mu+1}
=q_{\mu}+ \eta \left[\mathcal{P}^{\mathcal{M}}q_{\mu}-r_{\nu+1}\right],
\end{equation} 
where $I$ denotes the identity operator, and $\eta$ can be interpreted as a pseudotimestep, determined from linear stability theory for the $\mathcal{P}^{\mathcal{M}}$ operator. In our experiments, we find that performing only one iteration yields the best overall performance.
\end{article}
\clearpage

% Copy/paste for multiples of each file type as needed.

% enter figures and tables below here: %%%%%%%
%

